\documentclass[twocolumn]{aastex631}

\usepackage{newtxtext,newtxmath}
\usepackage[T1]{fontenc}
\usepackage{ae,aecompl}

\usepackage{graphicx}	% Including figure files
\usepackage{amsmath}	% Advanced maths commands
\usepackage{amssymb}	% Extra maths symbols

\usepackage[normalem]{ulem}
\usepackage{ragged2e}
\usepackage{xcolor}
\usepackage{color}

\def\PGPU{$\varphi-$GPU }

\begin{document}

%\title[]{Inward Bound: the incredible journey of massive black holes as they pair and merge -- II. Intermediate mass black holes in nucleated dwarf galaxies}

%\title[IMBH binaries in non-nucleated dwarf galaxies]{Stalling of intermediate mass black hole binaries at sub-parsec separation in non-nucleated dwarf galaxies}

%\title{Return of the Final Parsec Problem? Intermediate mass black hole binary dynamics in low density dwarf galaxies}

%\title{Navigating the final parsec for intermediate mass black hole binaries in low density dwarf galaxies}

\title{The potential for long-lived intermediate mass black hole binaries in the lowest density dwarf galaxies}

\correspondingauthor{Fazeel Mahmood Khan}
\email{fmk5060@nyu.edu, khanfazeel.ist@gmail.com}

\author[0000-0002-5707-4268]{Fazeel Mahmood Khan}
\affiliation{New York University Abu Dhabi, PO Box 129188, Abu Dhabi, United Arab Emirates}
\affiliation{Center for Astrophysics and Space Science (CASS), New York University Abu Dhabi}
\affiliation{Space and Astrophysics Research Lab (SARL), National Centre of GIS and Space Applications (NCGSA), Islamabad 44000, Pakistan}

\author{Fiza Javed}
\affiliation{Space and Astrophysics Research Lab (SARL), National Centre of GIS and Space Applications (NCGSA), Islamabad 44000, Pakistan}

\author{Kelly Holley-Bockelmann}
\affiliation{Department of Physics and Astronomy, Vanderbilt University, Nashville, TN 37240, USA}
\affiliation{Department of Physics, Fisk University, Nashville, TN 37208, USA}

\author[0000-0002-7078-2074]{Lucio Mayer}
\affiliation{Institut für  Astrophysik, Universität Zürich, Winterthurerstrasse 190, 8044, Zürich, Switzerland}

\author[0000-0003-4176-152X]{Peter Berczik}
\affiliation{Nicolaus Copernicus Astronomical Centre Polish Academy of Sciences, ul. Bartycka 18, 00-716 Warsaw, Poland}
\affiliation{Konkoly Observatory, Research Centre for Astronomy and Earth Sciences, HUN-REN CSFK, MTA Centre of Excellence, Konkoly Thege Mikl\'os \'ut 15-17, 1121 Budapest, Hungary}
\affiliation{Main Astronomical Observatory, National Academy of Sciences of Ukraine, 27 Akademika Zabolotnoho St, 03143 Kyiv, Ukraine}

%\author{Saeeda Sajjad}
%\affiliation{Space and Astrophysics Research Lab (SARL), National Centre of GIS %and Space Applications (NCGSA), Islamabad 44000, Pakistan}

\author[0000-0002-8171-6507]{Andrea V. Macci\`o}
\affiliation{New York University Abu Dhabi, PO Box 129188, Abu Dhabi, United Arab Emirates}
\affiliation{Center for Astrophysics and Space Science (CASS), New York University Abu Dhabi}
\affiliation{Max Planck Institut f\"{u}r Astronomie, K\"{o}nigstuhl 17, D-69117 Heidelberg, Germany}



\begin{abstract}

Intermediate Mass Black Hole (IMBH) mergers with masses $10^4 - 10^6$ $M_{\odot}$ are expected to produce gravitational waves (GWs) detectable by the Laser Interferometer Space Antenna (LISA) with high signal to noise ratios out to redshift 20. IMBH mergers are expected to take place within dwarf galaxies, however, the dynamics, timescales, and effect on their hosts are largely unexplored. In a previous study, we examined how IMBHs would pair and merge within nucleated dwarf galaxies. IMBHs in nucleated hosts evolve very efficiently, forming a binary system and coalescing within a few hundred million years. Although the fraction of dwarf galaxies ($10^7$ M$_{\odot} \leq$ $M_{\star} \leq 10^{10}$ M$_{\odot}$) hosting nuclear star clusters is between 60-100\%, this fraction drops to 20-70\% for lower-mass dwarfs ($M_{\star}\approx 10^7$ M$_{\odot}$), with the largest drop in low-density environments. Here, we extend our previous study by performing direct $N-$body simulations to explore the dynamics and evolution of IMBHs within non-nucleated dwarf galaxies, under the assumption that IMBHs exist within these dwarfs. To our surprise, none of IMBHs in our simulation suite merge within a Hubble time, despite many attaining high eccentricities $e \sim 0.7-0.95$. We conclude that extremely low stellar density environments in the centers of non-nucleated dwarfs do not provide an ample supply of stars to interact with IMBHs binary resulting in its stalling, in spite of triaxiality and high eccentricity, common means to drive a binary to coalescence. Our findings underline the importance of considering all detailed host properties to predict IMBH merger rates for LISA.

\end{abstract}


	
	%%%%%%%%%%%%%%%%% BODY OF PAPER %%%%%%%%%%%%%%%%%%

\section{Introduction}\label{sec-intro}

%\subsection{Massive black holes}

Most massive galaxies, including our Milky Way, harbor massive central black holes (MHBs) from $10^6 M_{\odot} - 10^9M_{\odot}$ \citep{Ferrarese+05,Kormendy+13,graham+15}. These massive black holes formed early during the evolutionary history of universe, as is evident from the presence of quasars and are observed in distant galaxies out to redshift $z \sim 10$ \citep{mia23a,mia23b,lar23}. Host galaxy properties, such as mass, velocity dispersion, and luminosity, are correlated with black hole mass \citep{har04,Gultekin+09, mcconnell+13,Davis2023}, suggesting a deep tie between the growth and evolution of both. Extrapolating these correlations suggests that lower mass galaxies should host BHs in the range $M_{BH} \simeq 10^4 M_{\odot} - 10^6 \, M_{\odot}$. Detecting black holes in this mass regime is observationally challenging, in part because the sphere of influence - a region where the MBH dictates the dynamics of the visible stars and gas, is below the resolution limit of space telescopes for all but the nearest galaxies. 

However with the improvement in the sensitivity of observational facilities over the last decade, there is substantial observational evidence that, like their massive counterparts, nearby galaxies with masses $M_{gal} \leq 10^{9.5} M_{\odot}$ do harbor central black holes with masses $M_{BH} \simeq 10^4 M_{\odot} - 10^6M_{\odot}$ \citep[e.g.][]{Reines_2015,Mezcua_2017,ngu19,Greene_2020, Majo21}, commonly referred to as Intermediate Mass Black Holes (IMBHs). IMBHs in dwarf galaxies are most easily detected via their AGN-like behaviour through, e.g. Doppler-broadened emission lines, X-ray or Radio brightness, or variability -- with the significant caveat that star formation, stellar binaries, and supernovae feedback can mimic an IMBH signal as well~\citep[see][for a review]{Askar23}. Still, infrared, optical, and x-ray searches for AGN-like signatures in the higher-mass end of dwarf galaxies reveal that the IMBH occupation fraction ranges from a few percent \citep{rei13,sat14,sat17, pol22} to more than forty percent~\citep{lem15}. These fractions could be regarded as a lower limit, because not all IMBHs in dwarfs may be accreting.
%Indeed a higher occupation fraction of $\> 50 \%$ is reported in \citep{ngu19}, who inferred IMBH presence in nearby dwarf galaxies by observing dynamical signatures of quiescent black holes. {\bf KHB: I don't see this fraction in that paper.}

Possible IMBH formation scenarios include their formation through Population III stars and subsequent growth via mergers and gas accretion \citep{mau01}, direct collapse of a massive gas cloud or protogalactic disk into black hole \citep{loeb94,beg06,dunn18,lat22} and gravitational runaway collision of massive stars as they sink under the effect of dynamical friction towards the center of a star cluster \citep{por04}.

\subsection{The dwarf galaxy landscape}

The term `dwarf' was coined to refer to a galaxy with a total luminosity $\leq 5 \times 10^8 \, L_\odot$ because a luminosity this low, coupled with the surface brightness limits at the time, implied a physical size much smaller than the Milky Way. The advent of large surveys and dedicated observing campaigns has revealed new and sometimes puzzling insights into the present-day census of dwarf galaxies, particularly at the faintest frontier, where at least 15 ultra-faint dwarf galaxies have been discovered in the Local Volume since 2021~\citep[e.g.][]{Cerny21, Cerny23b, Mau20, Bell22, Sand22, McQuinn23a, McQuinn23b, McNanna23, Cantu21, Simon21}. Ultra-faint dwarfs are some of the most dark matter-dominated structures known and display chemical enrichment signatures that clearly flag them as galaxies, but the stellar mass ($\lesssim 10^5 \, M_\odot$), size (10s of pc), and surface brightness ($\mu_V \gtrsim 31$) make them easy to be mistaken for star clusters. We now know that dwarf galaxies span nearly 40 magnitudes in surface brightness and range in size from $10 - 10^4$ pc ~\citep[e.g.][]{Simon19}, and that the dwarf landscape contains not only the classical dwarfs such as the Large and Small Magellenic Clouds, but galaxies that could be ultra-compact or ultra-diffuse, dark matter-deficient~\citep{Guo2020} or dark matter-dominated, extremely gas-rich~\citep{Hunter12} or completely devoid of gas, actively star-forming or quiescent, large or small. The dwarf landscape will continue to evolve rapidly with LSST, Euclid, and Roman observations, but efforts at the faintest frontier center on discovery, and higher-order questions, such as the possibility that these dwarf galaxies host IMBHs, remain unanswered.
For the purposes of this paper, `dwarf' refers to classical dwarfs unless otherwise specified, which tend to be small, dark-matter dominated, and may or may not be gas-rich or star-forming.

Classical dwarf galaxies can broadly be divided into two classes based on whether or not they have a central Nuclear Star Cluster (NSC) - an incredibly dense concentration of stars located at the center of a galaxy. A wealth of work indicates that the NSC occupation fraction peaks between 60-100\% at stellar masses $M_{\star} \sim 10^9$ M$_{\odot}$ and steadily drops to 20-70\% for $M_{\star}\sim 10^7 \, M_{\odot}$, with the smallest nucleation fraction in low-density environments ~\citep{denBrok14,ordenesBriceno2018,Eigenthaler2018,Hoyer21}. \citet{san19b} found that the decline of the nucleation occupation distribution at lower masses can be described by $f_n \propto {\rm log}  (M_{\star})^{1/4}$, falling to zero at $M_{\star} = 10^6 \, M_\odot$. So far, there is a strong correlation between the existence of an IMBH within a dwarf and the existence of a NSC~\citep{Askar23}. There is no evidence that IMBHs exist in the non-nucleated dwarfs we model here, though without the high central density and deep potential, identifying inactive IMBHs within non-nucleated dwarfs is beyond the scope of current observational capabilities.

%Dwarf galaxies with stellar masses $M_{gal} \sim 10^9$ M$_{\odot}$ exhibit a high occurrence of NSCs, ranging from 60\% in the Local Volume~\citep{Hoyer21} to over 95\% in the Coma cluster~\citep{denBrok14}.  the nucleation fraction drops with lower stellar mass. can be as low as $28$ \% as has been observed in Next Generation Fornax Cluster Survey \citep{Eigenthaler2018,ordenesBriceno2018}.

\subsection{IMBH binaries}

 Zoom-in cosmological simulations suggest that MBH binaries are usually hosted by galaxies with stellar masses exceeding $10^9 M_{\odot}$~\citep{dp23}. However, like their massive counterparts, dwarf galaxies also undergo mergers, albeit less frequently during their evolutionary history compared to more massive galaxies.  Such mergers would inevitably amass IMBH pairs in a merged galaxy system. 
As in case of massive galaxies, an IMBH binary is also expected to evolve in three distinct phases \citep{begelman+80}: IMBHs approach the center of the merger remnant due to dynamical friction furnished by background stellar, gas and dark matter distributions, then form a tight binary which subsequently evolves due to $3-$body interactions with background stars or further gas drag, and eventually merge if the tight binary phase is efficient in ushering the binary into the gravitational wave dominated regime. 

Although the dynamics and evolution of Supermassive Black Hole (SMBH) binaries have been studied extensively \citep{khan+11,Gualandris+12,holley+15,vasiliev+15,rantala+18,kha20,gua22}, the fate of IMBHs after a dwarf galaxy merges is relatively unexplored. If IMBHs can efficiently pair and merge in dwarf galaxies, they would be a promising gravitational waves sources for the ESA/NASA space-based observatory Laser Interferometer Space Antenna (LISA) \citep{ama23}, an adopted gravitational wave mission set to launch in the early 2030's. IMBHs mergers would be detectable by LISA with signal-to-noise ratios in the hundreds even out to redshift 20~\citep{Colpi24}. LISA observations of IMBH mergers will provide important insights to the formation and evolution of seed SMBHs at cosmic dawn, will trace the assembly history of galaxies, and will also place important constraints on the host galaxy structure, kinematics and stellar content. 

 Using high-resolution direct N-body simulations to accurately track the dynamical friction and few-body stages, \citet{Khan_Holley-Bockelmann2021} investigated IMBH pair dynamics in models of observationally-motivated nearby dwarf nucleated galaxies and concluded that IMBHs pair and merge very efficiently -- in merely a few hundred million years -- thanks to the high stellar densities furnished by NCSs. It was also observed that the host galaxy rotation and the orbit of the binary with respect to the host galaxy significantly impacts the evolution of IMBH binary's orbital parameters. Likewise, IMBH merger timescales of less than a Gyr were also obtained by \citet{mukh23}, who performed $N-$body simulations of IMBH evolution in mergers of idealised collisionally-relaxed NSCs in the absence of a surrounding dwarf galaxy, though their timescales depend strongly on the IMBH mass ratio.  \citet{tam18} performed simulations between disky dwarf galaxy mergers to explore the impact of the dark matter profile on IMBH pairing, but due to force softening, they couldn't explore IMBH dynamics in a bound state. 

These high-resolution studies are expensive, and since they zoom into the central part of a single galaxy, cannot accurately map to a global merger rate. For this, we must turn to cosmological simulations, which can explore the large scale IMBH dynamics, but are limited due to resolution effects and cannot probe the physics below the $\sim$ kpc regime. IMBHs in cosmological simulations are known to shrink inefficiently and wander at kpc scale separations from the centers of their hosts~\citep{Micic11, bellovary19a,bellovary19b}, though with the inability to resolve the central structures of the host, such as an NSC, the true distribution of wandering IMBHs remains unclear. Until a fully self-consistent cosmological simulation of IMBH dynamics over 20 orders of magnitude in scale becomes feasible, the state of the art is to introduce the results of the high-resolution IMBH merger simulations as sub-grid inputs into cosmological simulations. For example, \citet{de23} recently estimated IMBH merger timescales in a suite of cosmological simulations using a recipe that evolves IMBHs in dwarf galaxies by taking into account both the orbital decay caused by dynamical friction and the subsequent hardening of the IMBH binary by stellar and gaseous environments as witnessed in independent high resolution studies of such processes. In their simulation suite, where according to the authors the resolution is appropriate, IMBHs mergers happen on timescales that range from 0.8-8 Gyr.  Similarly, post-processing of the ASTRID simulation~\citep{Ni22} down to redshift 3 finds that IMBH merger timescales range from 1-10 Gyr~\citep{Chen22}, again with the caveat that none of the cosmological simulations have been able to model hosts with the central densities consistent with NSC-embedded galaxies, the majority of the dwarf galaxy population. Using a semi-analytic approach that paired local scaling relations between SMBHs and their host galaxies together with a recipe developed by \citet{seskha+15}, \citet{bia19} estimated that a binary's lifetime in low mass galaxies can be longer than a Hubble time.

Without a central NSC, IMBH pair dynamics and merger timescales in dwarf galaxies are expected to be strikingly different. It is known that a binary must scatter of order its own mass in stars to pass into the gravitational wave regime, and since the central stellar densities of non-nucleated dwarfs are between $10^4 - 10^8$ times lower than their nucleated counterparts, we expect IMBH mergers to be slow. Our question is: how do the mitigating factors of realistic kinematics and galaxy shape affect the IMBH dynamics in this extremely low-density regime?

%\citet{de23} recently estimated IMBH merger timescales in suite of their cosmological simulations using a recipe that evolves IMBHs in dwarf galaxies taking into account orbital decay caused by dynamical friction and subsequent hardening of IMBH binary by stellar and gaseous environments witnessed in independent high resolution studies of such processes. In their simulations suite, where according to the authors resolution is appropriate, IMBHs mergers happen in timescales that range from 0.8-8 Gyr.  

%\subsection{Our study}

We address this question by performing direct $N-$body simulations to study IMBH pair evolution in observationally-guided non-nucleated dwarf galaxy models. We construct non-nucleated dwarf galaxies based on observed scaling relations such as those between galaxy mass and effective radius, Sersic index and IMBH mass \citep{Reines_2015,Eigenthaler2018,ordenesBriceno2018,Greene_2020}.

This paper is arranged as follows: section 2 covers our galaxy models, simulation set-up and methods, section 3 presents the results, and section 4 concludes with a discussion of caveats, broader implications, and future work. 

\section{Initial Dwarf Galaxy Models and Numerical Techniques} \label{sec:galaxies} 
%%%%%%%%

%{\bf -- also, Krystal could remake the figures if you want to make them python-pretty :) }

Our suite of numerical simulations consists of five galaxy models covering a range of observed non-nucleated dwarf galaxies in the nearby Fornax and Virgo clusters. In particular, we build our models using the results of the Next Generation Fornax Survey (NGFS)~\citep{Munoz15,Eigenthaler2018,ordenesBriceno2018}. The masses of dwarf galaxies in the Fornax cluster are between 9.5 $\geq$ log M$_{\star}/$M$_{\odot}$ $\geq$ 5.5 \citep{Eigenthaler2018}, and we adopt $10^8$ M$_{\odot}$ for the stellar mass.
%; for this galaxy mass, the non-nucleated fraction in Fornax is 15\%~\citep{Munoz15}. 
The range observed for the Sérsic index $n$ for the dwarf galaxies in the Fornax Cluster is 0 $\leq$ $n$ $\leq$ 2, having a mean value of $n = 0.81$. We varied $n$ from 0.8 to 2.0 in our suite of galaxy models. Note that the Fornax galaxies we are using as reference are dwarf ellipticals, hence devoid of gas. The mean effective radius of non-nucleated dwarf galaxies in the Fornax cluster sample is $R_{\rm eff} = 0.56$ kpc, but can range from $0.18-2.22$ kpc \citep{ordenesBriceno2018}. We varied $R_{\rm eff}$ between $0.2-0.56$ kpc; our choice biases towards more compact non-nucleated dwarfs. The average axis ratio for this galaxy sample is observed to be $c/a=0.7$, and we adopt $b/a =0.9$. In the SMBH binary regime, triaxiality is key in ushering the binary through the scattering phase \citep{khan+11,preto+11,Gualandris+12,khan+13,Baile}. None of the Fornax dwarf galaxies have evidence of an IMBH; nevertheless, we introduce a primary IMBH of $1.2 \times 10^5$ M$_{\odot}$ at the center of each of our models using a scaling relation between galaxy stellar mass and IMBH mass \citep{Reines_2015,Greene_2020}:
\begin{equation}
    {\rm log} (M_{\rm IMBH}/M_{\odot}) = \alpha + \beta \,  {\rm log}(M_{\star}/10^{11} M_{\odot}),
\end{equation}\label{mimbh}
with $\alpha = 7.45$ and $\beta = 1.05$.

Table \ref{tab:resultsim1} summarizes the key parameters of our Fornax-inspired non-nucleated dwarf galaxy models. We use the publicly available software AGAMA \citep{vasi19} to generate our IMBH-embedded galaxy models in dynamical equilibrium using the Schwarzchild method.  We simulate each model with 2 million particles, for a particle mass resolution of 50 $M_\odot$. 
 
%%%%%%%%
%\begin{table*}
%\begin{center}
%\vspace{-0.5pt}
%\caption{Simulation Parameters} 
%\begin{tabular}{c c c c c c c}
%\hline
%Galaxy Model & $N$ & Mass ($10^8$ M$_{\odot}$) & $R_{\rm eff}$ (pc) & n  %& M$_{\rm IMBH}$ ($10^5$ M$_{\odot}$) & ${\rm b/a,c/a}$ \\
%\hline
%D0.8 & $2$ & $1.0$ & $580$ & $0.8$ & $1.2$ & $0.9 , 0.7$\\
%D1.0 & $2$ & $1.0$ & $560$ & $1.0$ & $1.2$ & $0.9 , 0.7$\\
%D1.5 & $2$ & $1.0$ & $560$ & $1.5$ & $1.2$ & $0.9 , 0.7$\\
%D1.5c & $2$ & $1.0$ & $200$ & $1.5$ & $1.2$ & $0.9 , 0.7$\\
%D2.0 & $2$ & $1.0$ & $200$ & $2.0$ & $1.2$ & $0.9 , 0.7$\\
%\hline
%\end{tabular}

%\vspace{15pt}
%\tablecomments{Column~1: Non-nucleated dwarf galaxy model. Here, D stands for dwarf and the number stands for the Sersic index. The c after D1.5c denotes a compact model that adopts the minimum observed $R_{\rm eff}$. Column~2: Number of particles in millions. Column~3: Galaxy model mass. Column~4: Galaxy model effective radius. Column~5: Sersic index. Column~6: Central IMBH mass. Column~7: intermediate to major ($b/a$) and minor to major ($c/a$) axes ratio. {\bf take out redundant columns, and I %think we can just combine with table 2}} \AM{AM: agreed, most numbers are %the same for all runs/galaxies, we can combine in a single table easier %to read} 
%\label{tab:simparam}
%\vspace{15pt}
%\end{center}
%\end{table*}
%%%%%%%%

\begin{figure}
\centerline{
  \resizebox{0.95\hsize}{!}{\includegraphics[angle=0]{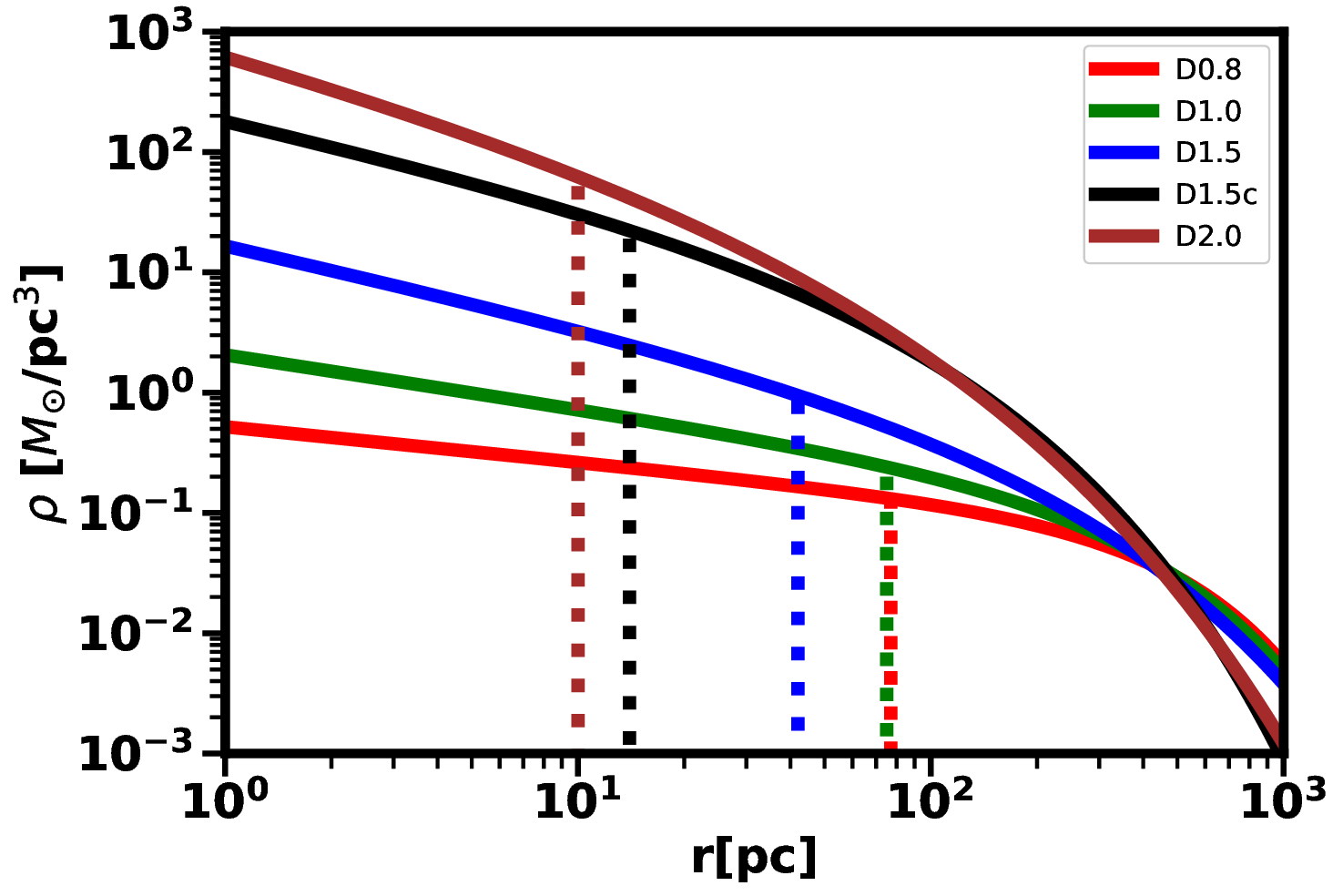}}
  }
\caption{
Stellar mass density profile of our dwarf galaxy sample as described in table \ref{tab:resultsim1}. The vertical dashed lines show the 3-d stellar density at the IMBH influence radius for the corresponding model. Our models span roughly $4$ orders of magnitude in central density, and the densest model is still several orders of magnitude less dense than nucleated dwarfs. 
} \label{fig:modeldens}
\end{figure}

Figure \ref{fig:modeldens} shows the initial mass density profile of our galaxy models within 1 kpc. Note that the central density within a parsec varies from below $0.5$ M$_{\odot}/{\rm pc}^3$ for D0.8 to a few few hundred  M$_{\odot}/{\rm pc}^3$ for D2.0. We tested all our models for stability by evolving them in isolation for $\sim$ 50 Myrs using the direct $N-$body code \PGPU\footnote{$N$-body code \PGPU: ~\url{ https://github.com/berczik/phi-GPU-mole}} \citep{berczik+11,Berczik2013} on the Vanderbilt ACCRE cluster. \PGPU integrates stellar and IMBH orbits using a $4^{th}$-order Hermite integrator with the option of softening. We employ $\epsilon_{IMBH} = 0$ softening for IMBH-IMBH interactions and a softening of $\epsilon_{\star,\star}=10^{-2}$ parsec (pc) for star-star interactions. For IMBH-star interactions, an initial mixed softening parameter is calculated using equation \ref{eq:soft}.
\begin{equation}
\epsilon_{\rm{\star,IMBH}} = \varepsilon_{\rm{corr}} \cdot \sqrt{\frac{\epsilon_{\star}^2 + \epsilon_{\rm IMBH}^2}{2}}.
\end{equation} \label{eq:soft}
Here we use the correction factor $\varepsilon_{\rm{corr}} = 0.1$ resulting in softening for the star-IMBH interactions to be $\simeq$ $0.007$ pc. Once the IMBH binary is bound, we reduce 
$\varepsilon_{\rm{corr}}$ to 0.001 for a resulting softening $\epsilon_{\rm{\star,IMBH}}$ to $7 \times 10^{-5}$ pc. The present code is well-tested and already used in earlier large scale studies~\citep{2014ApJ...792..137Z, 2018ApJ...868...97K, 2014ApJ...780..164W, Ber2022, AB2019}. For more details on generating Sérsic models as well as the \PGPU code, we refer the reader to section 2 of \citet{Khan_Holley-Bockelmann2021}.

%%%%%%%%
%%%%%%%%

%%%%%%%%
%\begin{figure}
%\centerline{
%  \resizebox{0.95\hsize}{!}{\includegraphics[angle=0]{sigmarel.eps}}
%  }
%\caption{
%Massive Black Hole mass and velocity dispersion relation from .......  {\bf KHB: do we need this as a figure?}
%} \label{fig:massdispersion}
%\end{figure}
%%%%%%%%

%%%%%%%%
%\subsection{Initial orbits} \label{subsec:iniorb}

A secondary IMBH four times less massive than the primary was introduced on an elliptical orbit with an eccentricity of 0.5 at an apocenter distance of 100 pc. This initial position of the secondary IMBH is well outside the influence radius of the primary IMBH ($r_{\rm infl} \sim 10-60$ pc), see table \ref{tab:resultsim1}. Although we do not simulate the merger of the progenitor IMBH-embedded dwarf galaxies, the choice of this starting
distance is comparable, within factors of a few, to the spatial resolution of zoom-in cosmological simulations and isolated simulations of IMBH-embedded dwarf galaxy mergers, so in a sense our simulations tend to pick up where the larger-scale ones leave off.

%\lipsum
\begin{figure*}
    \centering{
    \resizebox{0.99\hsize}{!}{\includegraphics[angle=0]{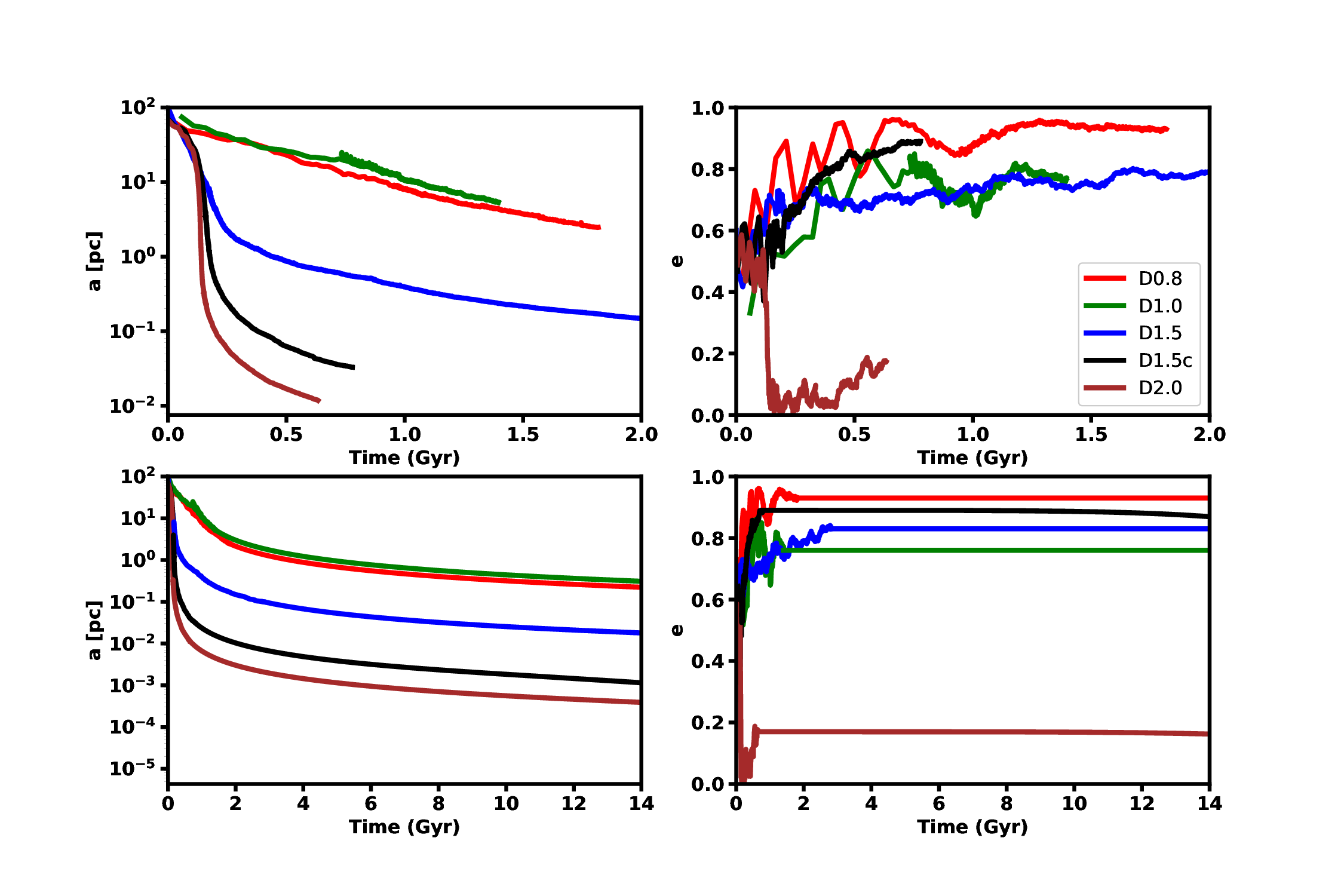}}
    }
    \caption{IMBH pair and binary evolution in our non-nucleated dwarf galaxy models. Top left: IMBH orbit semi-major axis. Top right: IMBH orbit eccentricity evolution. Bottom left: Combined semi-major axis evolution from our $N$-body run and semi-analytic estimates for the IMBH binary. Bottom right: Combined eccentricity evolution from our $N$-body run and semi-analytic estimates for the IMBH binary. }
    \label{fig:img1}
\end{figure*}
%\lipsum

\section{Results}\label{sec:results}

%\subsection{Binary evolution} \label{subsec:reciepe}
Following \citet{mer06R}, we define the influence radius ($r_h$) of IMBH binary to be the radius of the sphere that encloses a stellar mass twice that of the binary; there are 5000 particles within the sphere of the influence of our models. A massive binary is called a {\sl hard binary} when its semimajor axis is equal to:

\begin{equation}
    a_h = \frac{r_h}{4}\frac{q}{(1+q)^2},
\end{equation} \label{ah}

\noindent where $q = M_2/M_1$, $M_1$ is the mass of the primary IMBH and $M_2$ is the mass of the secondary IMBH. For our case, $a_h = 0.04 \times r_h$.

The IMBH pair evolution for our runs is shown in figure \ref{fig:img1}. The top panels feature the semi-major axis and eccentricity evolution directly from our $N-$body simulations, respectively. Note that we described the eccentricity when the IMBH pair is unbound using equation \ref{eq:eapoperi}, and when the IMBH becomes a hard binary, we calculate semi-major axis $a$ and eccentricity $e$ for a Keplerian orbit (equations \ref{eq:a} and \ref{eq:eccn-kepler}, respectively).

\begin{equation}
    e=\frac{r_a - r_p}{r_a+r_p},
    \label{eq:eapoperi}
    \end{equation}
where $r_a$ and $r_p$ are the apo and pericenter distances of a binary orbit in relative coordinates. For a bound binary, the semi-major axis is:

\begin{equation}
    a = -\frac{1}{2}\frac{M_{\bullet}}{E_{\rm IMBH}},
    \newline
    \mathrm{ where~}
    \newline
    E_{\rm IMBH} = -\frac{M_{\bullet}}{r} + \frac{1}{2}v^2,
    \label{eq:a}
    \end{equation}
    is the specific orbital energy, $M_{\bullet}$ is the IMBH binary mass, $r$ is relative separation and $v$ is the relative velocity of the black holes.

The Keplerian eccentricity is expressed by:
\begin{equation}
        e=\sqrt{1+\frac{2E_{\rm IMBH} h^2}{\mu^2}},
        \label{eq:eccn-kepler}
    \end{equation} \label{kepe}
    \noindent where h is the specific relative angular momentum and $\mu$ = G $\times M_{\bullet}$.

We study the evolution of IMBH binaries until the late phase of the hard binary regime, where the binary's dynamics is determined by the scattering of individual stars with the massive binary through few-body interactions. We obtain the IMBH binary hardening rate for each run by fitting a straight line to the inverse semi-major axis during last one hundred million years of binary evolution in our run; $s = d/dt(1/a)$. Table \ref{tab:resultsim1} shows the hardening rates for all runs. As expected, it is clear that $s$ increases with central stellar density (figure \ref{fig:modeldens}), i.e. models with higher Sérsic indicies and smaller effective radii. Denser galactic nuclei have a larger pool of stars in the vicinity of the binary to interact with and efficiently lose energy in the process \citep{khan+12a,vasi13,gua22}.  

%%%%%%%%
\begin{table*}
\begin{center}
\vspace{-0.5pt}
\caption{IMBH Binary and Host Parameters and Merger Results} 
\begin{tabular}{l c c c c c c c c c c}
\hline
Run & $R_{\rm eff}$ (pc) & $\rho_{0}$ (M$_{\odot}/{\rm pc^3}$)  & $r_{h}$(pc)  & $\rho_{r_h}(M_{\odot}/{\rm pc^3})$ & $a_f/a_h$ & $s( {\rm pc^{-1}/Myr})$ & $e_{\rm f}$ & $T_{\rm coal} ({\rm Gyr})$ & $T_{\rm coal} (e_{0.95})$ & $T_{\rm coal} (e_{0.99})$\\
\hline
D0.8 & $560$ & $4.5 \times 10^{-1}$ & $75$ & $1.7 \times 10^{-1}$ & $ 0.8 $& $0.00025$ & $0.93$ & No & $600$ & $224$ \\
D1.0 & $560$ & $1.7 \times 10^{0}$ & $72$ & $1.9 \times 10^{-1}$ & $1.15 $&  $0.0003$ & $0.76$ & No & $500$ & $180$ \\
D1.5 & $560$ & $1.8 \times 10^{1}$ & $40$ & $1.3 \times 10^0$  & $ 0.06 $ & $0.004$ & $0.83$ & No & $70$ & $24$ \\
D1.5c & $200$ & $1.9 \times 10^{2}$ & $14$ & $2.6 \times 10^1$  & $ 0.036 $ & $0.055$ & $0.92$ & No & $13$ & $3.2$ \\
D2.0 & $200$ & $5.7 \times 10^{2}$ & $10$ & $9.1 \times 10^1$ &  $0.017 $ & $0.2$ & $0.17$ & No & $3.45$ & $1.3$ \\
\hline
\end{tabular}\label{tab:resultsim1}
\vspace{15pt}

\tablecomments{Column~1: Non-nucleated dwarf galaxy model. Here, D stands for dwarf and the number stands for the Sersic index. The c after D1.5c denotes a compact model that adopts the minimum observed $R_{\rm eff}$. Column~2: Galaxy model effective radius. Column~3: Stellar density within the inner parsec. Column~4: Influence radius for a $1.5 \times 10^5 M_{\odot}$ IMBH binary (initial model). Column~5: Stellar density at the influence radius (initial model). Column~6: Ratio of the final semi-major axis in our $N-$body run to that of hard binary semi-major axis.} Column~7: Stellar hardening rate $s$. Column~8: Final value of $e$ at the end of direct $N-$body run. Column~9: IMBH binary merger time. Column~10 and 11: IMBH merger times for assumed values of $e = 0.95$ and $e = 0.99$ respectively.
\end{center}
\vspace{15pt}
\end{table*}
%%%%%%%%

%%%%%%%%
\begin{figure}
\centerline{
  \resizebox{0.95\hsize}{!}{\includegraphics[angle=0]{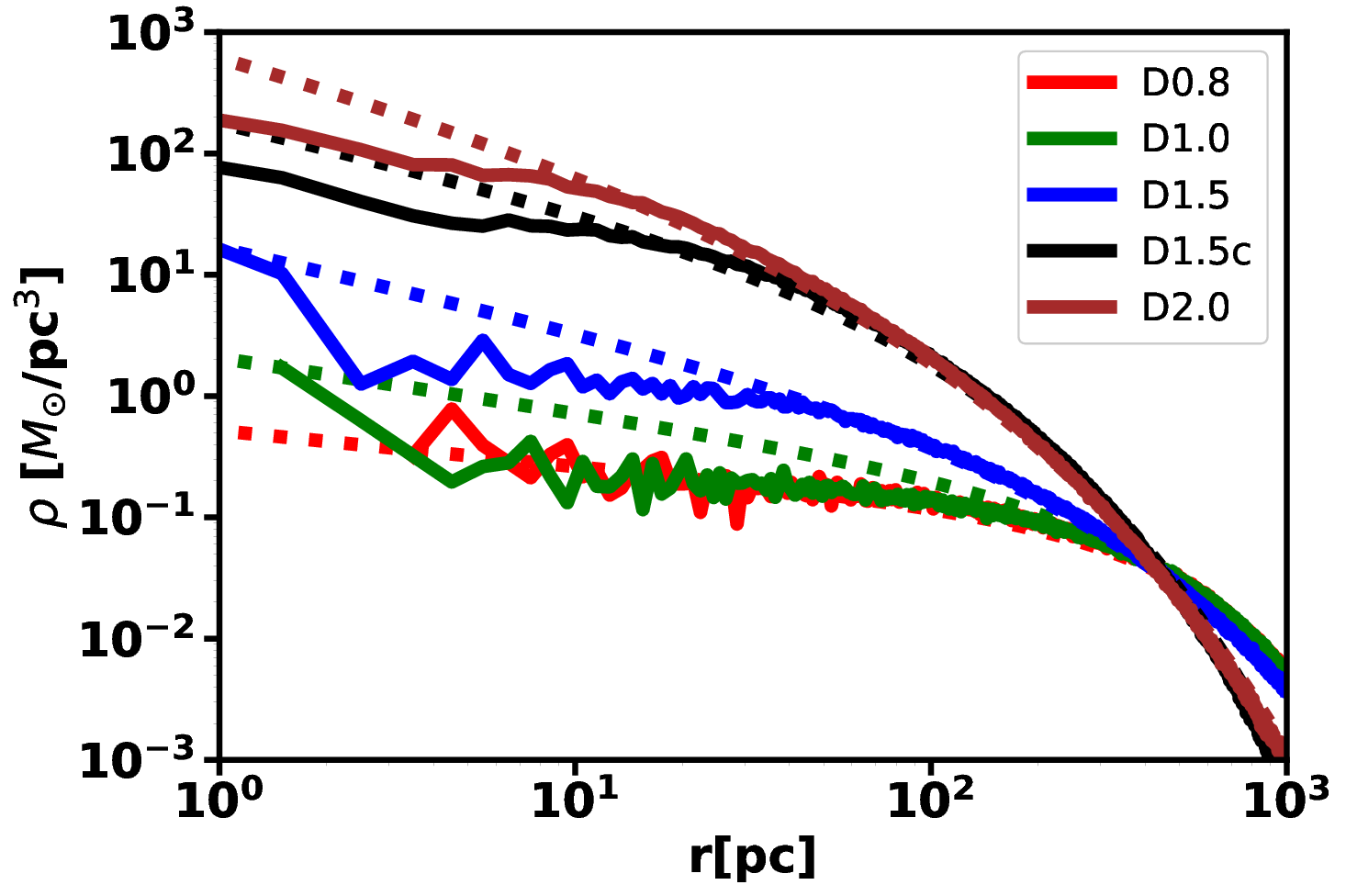}}
  }
\caption{
Mass density profile of our dwarf galaxy sample at the end of simulation run as described in table \ref{tab:resultsim1}. Central densities have dropped marginally for low Sersic index models and substantially for high $n$ and smaller effective radius models. For comparison, we have also plotted the initial profiles (dotted lines).
} \label{fig:final-dens}
\end{figure}
%%%%%%%%

After we stop the simulation, we estimate the subsequent evolution of the IMBH binaries over the course of $14$ Gyrs by fixing the hardening rate $s$, which reflects the rate at which stellar encounters extract energy from the IMBH binary, to the final value. We also take into account the expected loss of orbital energy due to gravitational wave emission, as described by \cite{peters+63}. We assume that the final eccentricity remains constant after the end of our simulations. Neither the hardening rate nor the eccentricity truly remain constant. However, we can derive a lower limit to the IMBHs merger time by assuming the eccentricity grows to 0.99 after the simulation ends. 

IMBH binary evolution for models D0.8 and D1.0 exhibit similar initial behaviour: the semi-major axis shrinks slowly from $100$ pc to a few parsec in roughly $1.5$ Gyr. This analogous time evolution is due to the very similar density profiles (see fig. \ref{fig:modeldens}), resulting in a nearly identical dynamical friction time. Despite their high eccentricity ($0.76$ and $0.93$, respectively), the IMBH binary in these models stall at about a parsec separation during its 14 Gyr of evolution due to low hardening rates ($0.00025$ $\&$ $0.0003$  ${\rm pc^{-1}/Myr})$. These models are characterized by Sérsic indices and effective radii consistent with what is observed 
%AM that are similar to the most likely values of these parameters observed 
in non-nucleated dwarf galaxies. Therefore, if non-nucleated dwarf galaxies contained IMBHs and merged with another non-nucleated dwarf galaxy, the IMBHs would likely remain as a bound binary throughout cosmic time. 
%Gas accretion or stellar tidal disruption could hypothetically render either or both IMBHs electromagnetically bright, with characteristic doppler boosting evident in a light curve or spectrum~\citep{Charisi22}. Figure \ref{fig:imbhvel} shows the doppler motion of the IMBHs in the D0.8 and D1.0 models, with the secondary orbiting at velocities as high as 70 and 30 km/sec, respectively.   
%{\bf KHB what is the orbital period? -- 0.92 Myr for D0.8 and 1.5 Myr for D1.0}

 %%%%%%%%
%\begin{figure}
%\centerline{
%  \resizebox{0.95\hsize}{!}{\includegraphics[angle=0]{velocities-D01-02.eps}}
%  }
%\caption{
%Velocities of IMBHs in binary center of mass frame of reference for models D0.8 and D1.0 at the time when we stop the respective simulations. The orbital period of the binary is 0.92 Myr for D0.8 and 1.5 Myr for D1.0, respectively. 
%} \label{fig:imbhvel}
%\end{figure}
%%%%%%%%
 
 For model D1.5, we maintain the effective radius while raising $n$ to $1.5$, resulting in an approximately tenfold increase in central density. As a consequence, the IMBH binaries evolve with a higher hardening rate of $s=0.004$ ${\rm pc^{-1}/Myr})$. Again, the eccentricity remains high $e \sim 0.83$, but despite a higher hardening rate and high eccentricity, the IMBHs in binaries do not merge within a Hubble time. However, they do manage to shrink to a smaller separation of approximately $0.01$ pc.
 
Finally, our last two models, namely D1.5c and D2.0, both exhibit an effective radius of $200$ pc. These models are on extreme end of non-nucleated dwarf galaxies in terms of their parameter choices and hence may represent a tiny fraction of an already small population of dwarfs. D1.5c and D2.0 have higher central stellar densities, $200$ M$_\odot$/pc$^3$, and $600$ M$_\odot$/pc$^3$, respectively -- placing them among the densest non-nucleated dwarf galactic nuclei. This high density translates into stronger hardening, $s=0.055$ and $s=0.2$ ${\rm pc^{-1}/Myr})$ for models D1.5c and D2.0, respectively. The IMBH binary in model D1.5c again attains a high value of $e \sim 0.92$, whereas model D2.0 hosts a binary with a small $e \sim 0.17$. This lower eccentricity for D2.0 is consistent with earlier findings that SMBH binaries evolving in models with cuspy inner density profiles form and maintain low eccentricities in the hard binary regime \citep{khan+12a}. It takes roughly 14 Gyr for the IMBH binary semi-major axis to shrink below a milliparsec separation, at which point significant gravitational wave emission still must occur before the IMBHs can merge. 
%still orders of magnitude far from Schwarzchild radius ($\sim 10^{-8}$) pc, where the eventual merger of IMBHs happen.   {\bf yes, but do they shrink to the GW regime? Not R$_s$.}
We observe high eccentricities for IMBH binaries in our models, with $e \sim 0.7-0.95$, except for model D2.0, which has an eccentricity value of 0.17. This contrasts with the low or intermediate eccentricity values reported by \citet{ogi20} and \citet{mukh23}. The contrasting behavior is due to the fact that these authors studied IMBH binary evolution in mergers of nuclear star clusters, which have very high central densities. As shown in earlier studies by \citet{khan+12a}, \citet{khan18a} and \citet{gua22}, dense and cuspy galaxy models tend to host massive black hole binaries with low eccentricities, whereas those with shallower cusps or lower central densities tend to host binaries with higher eccentricities. Indeed, in our model D2.0, which has a higher density compared to the other models, the IMBH binary forms and retains a low eccentricity.

We tested a hypothetical scenario in which IMBH binaries achieve very high eccentricities either through stellar scattering, counter-rotation \citep{sesana11,holley+15,Mirza+17}, or via the Kozai-Lidov mechanism \citep{kozai62,lidov62}. We estimate the merger time for two hypothetical eccentricities, namely, $e = 0.95$ and $e = 0.99$ and the results are shown in columns $9$ and $10$ of table \ref{tab:resultsim1}. The IMBH binaries in models D01, D02 and D03 do not achieve coalescence in a Hubble time even with these high eccentricities, but IMBHs in model D04 and D05 do merge in a few Gyrs. As stated earlier, the higher stellar densities seen in models D04 and D05 are outliers and do not represent the peak population of non-nucleated dwarfs in nearby clusters.

%%%%%%%
%\begin{figure*}
%\centerline {
%\resizebox{0.95\hsize}{!}{\includegraphics[angle=270]{ngc5102bh.ps}}
%  }
%\caption{
%IMBH evolution for NGC 5102, as in figure \ref{fig:m32multi}. 
%} \label{fig:ngc5102param}
%\end{figure*}
%%%%%%%
\section{IMBH Merger Estimates}

\begin{figure}
\centerline{
  \resizebox{0.95\hsize}{!}{\includegraphics[angle=0]{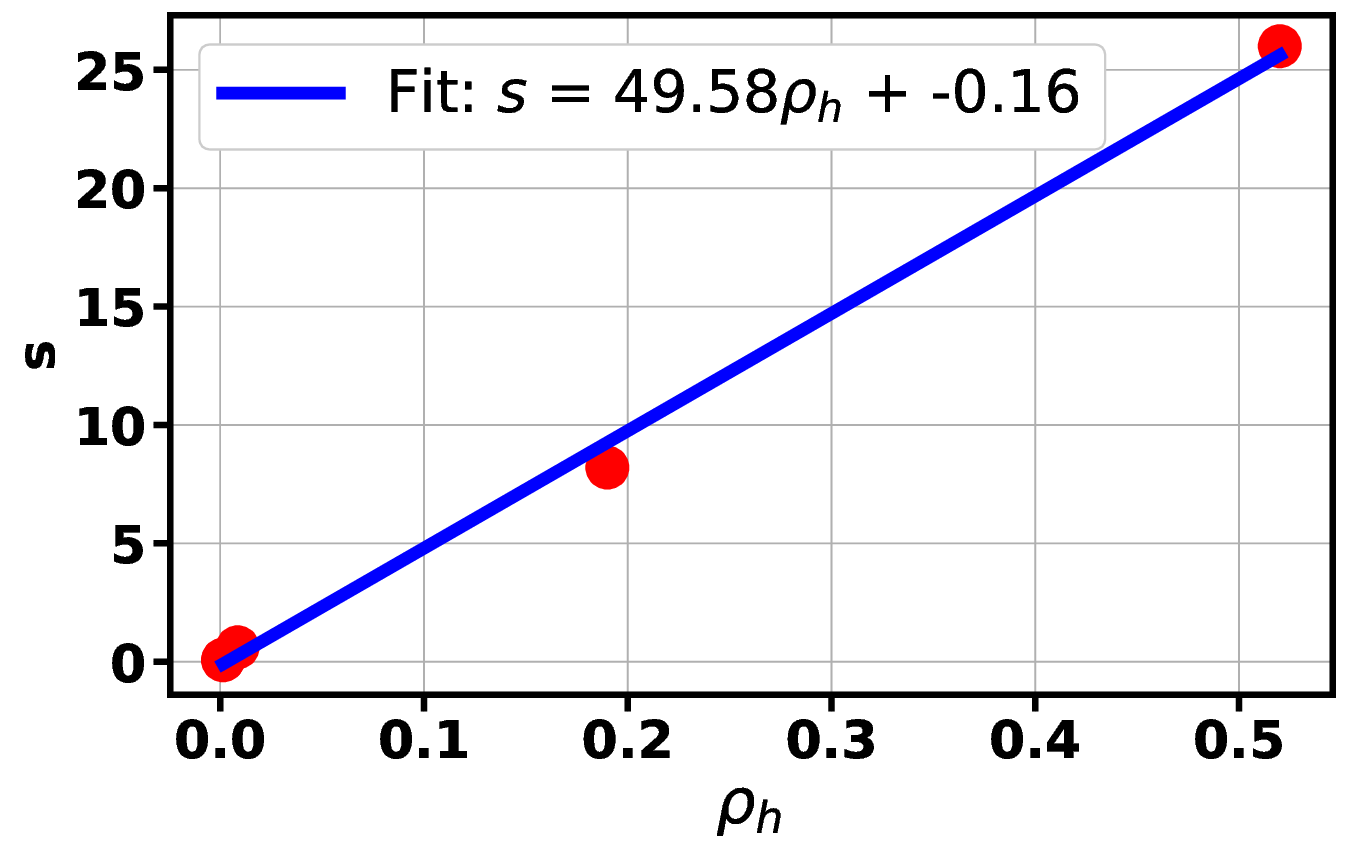}}
  }
\caption{
IMBH binary hardening rates (model units) vs density (model units) at binary influence radius. The straight line shows the linear fit to the data. 
} \label{fig:srhofit}
\end{figure}

We used the IMBH binary and host galaxy parameters from our simulation suite to estimate the stellar density at the binary influence radius, $\rho_h$, for which an IMBH binary can merge within a Hubble time for non-nucleated dwarf galaxies. We plot hardening rate, $s$ vs $\rho_h$ in figure \ref{fig:srhofit}, and apply a linear fit to the data, with the fit parameters indicated in the figure. 

We use this correlation to estimate the IMBH binary hardening rate $s_{ST}$ in stellar hardening regime and equate it to the hardening rate $s_{GW}$ from GW emission \citep{peters+63} to obtain the semi-major axis value $a_{GW}$ where the transition from stellar dominated hardening to GW dominated hardening occurs. Subsequently, we estimate IMBH binary lifetime using:

%As in our earlier studies, we estimate the binary lifetime using

\begin{equation}
    t(a_{gw}) = \frac{1}{s*a_{GW}}
\end{equation}

In our earlier studies \citep{khan+12a, seskha+15}, we have shown that this approach works reasonably well, and the estimated binary lifetime is accurate to within about 10\% of semi-analytic estimates using the combined effects of stellar hardening and gravitational wave emission from a hard binary's semi-major axis value for the complete evolution. 
%As IMBH binaries spend most of their lifetime in transition regime, the absence of considering stellar hardening in addition to GW hardening in our calculations compensates the earlier phase of transitioning from a hard binary semi major axis value to $a_GW$. 
At a specified density value $\rho_h$ and eccentricity $e$, an IMBH binary lifetime can be approximated:
\begin{equation}
t\left(\text Gyr\right) = \left(\frac{1.15 \times 10^{12}}{\left(\frac{M_1}{10^5 M_{\odot}}\right) \cdot \left(\frac{M_2}{10^5 M_{\odot}}\right) \cdot \left(\frac{M_{\bullet}}{10^5 M_{\odot}}\right) \cdot F(e)} \cdot \left(\frac{\rho_h}{2}-0.16\right)^{-4}\right)^{1/5},
\end{equation} \label{tofrho}
where $F(e)$ =$\left(1 - e^2\right)^{-3.5} \left(1 + \frac{73}{24} e^2 + \frac{37}{96} e^4\right)$.
From equation 8,  the density value at influence radius required for a given eccentricity of the IMBH binary such that it achieves coalescence in a Hubble time is given by   

%\begin{equation}

%\rho = \frac{1}{49.59} \left( \left( \frac{1}{\frac{64 M_1 M_2 (M_1 + M_2) \left(1 - e^2\right)^{-3.5} \left(1 + \frac{73}{24} e^2 + \frac{37}{96} e^4\right)}{5 c^5} \cdot t^5} \right)^{\frac{-1}{4}} + 0.16 \right)

%\end{equation}

%\begin{equation}
%\rho = \frac{1}{49.59}  \left( \left({\frac{64 M_1 M_2 (M_1+M_2) F(e)}{5 c^5} \cdot t^5} \right)^{-1/4} + 0.16 \right),
%\end{equation}
%where $F(e)$ =$\left(1 - e^2\right)^{-3.5} \left(1 + \frac{73}{24} e^2 + \frac{37}{96} e^4\right)$, e is eccentricity, $M_1$ is the primary IMBH mass, and $M_2$ is the secondary IMBH mass.

\begin{equation}
\rho_h\left(\frac{M_{\odot}}{pc^3}\right) = 2  \left( \left({\frac{\left(\frac{M_1}{10^5 M_{\odot}}\right) \cdot \left(\frac{M_2}{10^5 M_{\odot}}\right) \cdot \left(\frac{M_{\bullet}}{10^5 M_{\odot}}\right) \cdot F(e)}{1.15 \times 10^{12}} \cdot \frac{t^5}{Gyr}} \right)^{-1/4}+0.16 \right).
\end{equation}\label{rhooft}

Note that this is only correct for non-nucleated dwarf galaxies and IMBH masses comparable to those being considered in the current study.
%In our model, the total mass of all galaxies is normalized to \textbf{1.0}, which corresponds to \( 10^8 \) solar masses (\( M_{\odot} \)). The length unit is set at \textbf{100} parsecs (\textbf{pc}), leading to a time unit of approximately \textbf{1.49} million years (\textbf{Myr}) and a value of \( c \) equal to \textbf{4571}. 
%In this context, \( t \) represents time in our model units, where the density \( \rho \) 
%facilitates the merger of intermediate-mass black hole (IMBH) binaries.

 %%%%%%%%
%%%%%%%%

%For a given value of eccentricity $e$, density at the influence radius to furnish the coalescence is plotted in figure \ref{fig:rhoe}. We choose a $t$ value that corresponds to 13 Gyr in physical units. This the time once a hard binary forms assuming that IMBH pair separation shrinks to a hard binary separation in about a Gyr. 
%For a circular IMBH binary to achieve a merger within a Hubble time, the density at the influence radius needs to be 100 $M_{\odot}/pc^3$. In contrast, for an IMBH binary with an orbital eccentricity of approximately 0.95, a density of 10 $M_{\odot}/pc^3$ is required at the influence radius. From figure \ref{fig:modeldens} and table \ref{tab:resultsim1},  we notice that two of our models namely, D1.5c and D2.0 have stellar densities in the range of 10-100 $M_{\odot}/pc^3$. Indeed from column 9 of table \ref{tab:resultsim1}, merger times of IMBH binaries in these models is shorter than a Hubble time for appropriate $e$ values. 

\begin{figure}
\centerline{
  \resizebox{0.95\hsize}{!}{\includegraphics[angle=0]{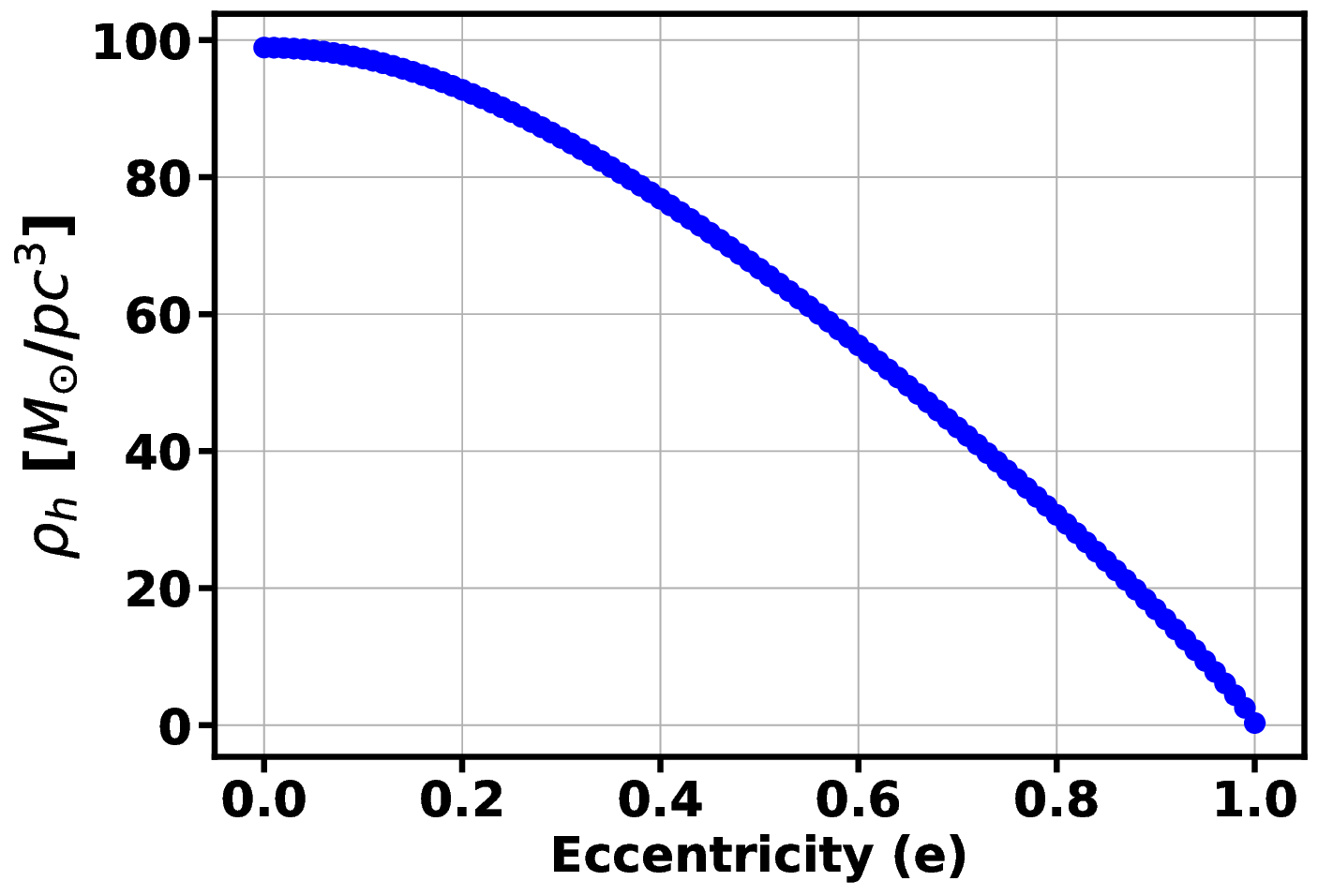}}
  }
\caption{
Central stellar density required for an IMBH binary to merge within a Hubble time, if the host is a non-nucleated dwarf galaxy, as a function of eccentricity. 
} \label{fig:rhoe}
\end{figure}

For a given eccentricity \( e \), the density at the influence radius required 
for IMBH binary coalescence in $t=\,13$ Gyr is illustrated in Figure~\ref{fig:rhoe}. We assume an additional billion year for the IMBHs go through the dynamical friction phase to become a hard binary. For a circular IMBH binary within a non-nucleated dwarf galaxy to merge within a Hubble time, the density at the 
influence radius must be \( 100 \, M_{\odot}/\text{pc}^3 \). In contrast, for an 
IMBH binary with an orbital eccentricity of approximately 0.95, a density of 
\( 10 \, M_{\odot}/\text{pc}^3 \) is required at the influence radius. As shown 
in Figure~\ref{fig:modeldens} and Table~\ref{tab:resultsim1}, two of our models, 
namely D1.5c and D2.0, have stellar densities ranging between \( 10 \) and 
\( 100 \, M_{\odot}/\text{pc}^3 \). Specifically, as indicated in column 10 of 
Table~\ref{tab:resultsim1}, the merger times of IMBH binaries in these models 
are shorter than a Hubble time for appropriate values of \( e \).

%%%%%%%%
%%%%%%%%
\section{Discussion and Conclusion}

In this study, we performed a suite of high-resolution direct $N-$body simulations to explore the dynamics of IMBHs in non-nucleated dwarf galaxies. IMBH mergers are a prime target for LISA because they are long-lived and bright sources; IMBHs merge at the peak sensitivity of the LISA band and since the inspiral would be observable by LISA for at least a year before merger, the waveform may encode astrophysical information about accretion physics that could otherwise be inaccessible~\citep{Garg22}.

The galaxy models were built using the observed structural and kinematic scaling relations of non-nucleated dwarfs in nearby Virgo and Fornax clusters. An IMBH with mass $1.2 \times 10^5$ M$_{\odot}$, again adopted from scaling relations of IMBH versus stellar mass, was placed at the center of each dwarf galaxy. A secondary IMBH of 1/4 the primary IMBH mass was placed on a moderately eccentric orbit at an initial separation of $100$ pc. After roughly 1 Gyr of evolution in our simulation suite, IMBH binaries end up around a separation that can range from $10$ pc- $10^{-2}$ pc, depending on the central density. We run our lower density models (D0.8,D1.0 and D1.5) for a longer duration, approximately 2 Gyr. Subsequently, we combine the $3-$body hardening rate with the energy loss by gravitational wave emission for each run, and evolve the binaries semi-analytically until 14 Gyr. None of the IMBH binaries in our suite merge within a Hubble time with the parameters that we witness in our runs. This is in line with \citet{bia19}, who estimated a timescale longer than a Hubble time for IMBH binaries, and stands in striking contrast to the IMBH binary merger timescales of well below 1 Gyr observed in nucleated dwarfs \citep{Khan_Holley-Bockelmann2021} or those recorded in mergers of NSCs \citep{ogi20, mukh23}. These contrasting outcomes of IMBH binary evolution is attributed to stellar densities that are 3-4 order of magnitudes lower than their nucleated counterparts. In non-nucleated dwarfs, IMBH binaries have a much smaller pool of stars to interact with, vital for ushering them into a gravitational wave-dominated regime.   
%
%{\bf KHB: this is not a given -- maybe this does happen. Maybe it happened and the non-nucleated ones have already had it happen, but won't in the future? Needs a re-write} Given that dwarf galaxies should assemble plenty of  IMBHs during their cosmic evolution either through mergers of stellar mass BHs and lower mass IMBHs ~\citep[e.g.][]{lupi2016,sas23}, or directly via stellar dynamical processes in clusters,  our study suggests that a large number of non-nucleated dwarfs may have stalled IMBH pairs with a separation around $1$pc -$1$mpc depending on their central densities and evolutionary history.  This population of stalled IMBH binaries is then predicted for separations much smaller
%than those of wandering IMBHs found in cosmological simulations in low mass galaxies~\citep[][e.g.]{Bell22} as well as in dwarf galaxy mergers
%with shallow dark matter profiles~\citep[][e.g.]{tam18}, which end up at separations from 100 pc to kiloparsecs. 

Our results disagree with \citet{de23}, who estimated IMBH mergers in suite of their cosmological simulations that have similar velocity dispersion and stellar density towards the center. The authors report IMBH mergers timescales in range $0.8-8$ Gyr, once a binary forms. The main reason for this discrepancy is the recipe the authors adopt to estimate the IMBH binary timescale in hardening phase. They use a formalism from \citet{vasiliev+15}, however, it should be noted that the parameters for hardening timescale presented in that study were extracted for galaxies that have cuspy density profiles with power law indices ranging from 1-2, observed in centers of bulges of late type spiral galaxies or intermediate mass elliptical galaxies. This setting is in contrast to low mass and low density shallow cusp/core dwarf ellipticals that are under discussion here. Our results suggest that it is dangerous to extrapolate results of cosmological simulations, that cannot resolve the hardening phase, and conclude that IMBHs in dwarfs will merge in the LISA band.

Taken on face value, our study suggests that some number of non-nucleated dwarfs may host stalled IMBH pairs with a separation around $1$pc -$1$mpc, depending on the central density. This population of stalled IMBHs would be on a much smaller scale than the wandering IMBHs of cosmological simulations~\citep{Bell22,bellovary19b, Micic11} or the IMBHS within dwarf galaxy mergers with shallow dark matter profiles~\citep{tam18}, which predict separations of 100s to 1000s of parsecs. Take care, though: the low-density dwarfs of the present day may have had a nucleated past, or at least a higher density one. It is even conceivable that an earlier IMBH merger could have sculpted a nucleated dwarf into the present low-density system by ejecting stars in the few-body scattering phase; compounding the conundrum, the IMBH could itself have been ejected after the merger via gravitational wave recoil. In this scenario, the non-nucleated population could archive a prior stage in which binary black holes merged efficiently. Distinguishing this dwarf galaxy formation mechanism from others, such as baryonic feedback~\citep{brooks14}, may require kinematic information for intracluster stars.

If non-nucleated dwarfs are in place at high redshift, then 
 the inability of IMBHs to merge within them could suggest that IMBH growth is inefficient in this dwarf galaxy type. Indeed,
had we simulated lower mass IMBHs our conclusions would be even stronger, as both the dynamical friction and 3-body hardening phases would become
longer.  With the merger channel for IMBH growth suppressed, in a gas-poor system, there would leave no option to grow
a sizable IMBH from lower mass black hole progenitors. This might explain why the majority of detected AGNs in dwarf galaxies are found in gas-rich disky systems with NSCs that resemble a small (pseudo)bulge~\citep{Kimbrell23}. 
%It is tempting to make an even more general inference, arguing that, since AGNs in dwarfs are, by definition, associated with the subset of IMBHs that are able to accrete gas efficiently, then in such systems efficient gas fueling to the center, which will inevitably cause the formation of a NSC or even a small bulge-like component in the dwarf, is a common occurrence. In such a scenario
%the trends found by \citet{Kimbrell23} would be naturally explained.
%As a caveat, we notice that AGNs, albeit in lower numbers, are found also in dwarf ellipticals
%(Kimbrell et al. 2021;2023), although there is no systematic study yet that compares their occurrence in nucleated vs. non-nucleated dwarf ellipticals.  There might well be an hidden large population of non-accreting IMBHs in dwarf ellipticals. 
Following this line of logic, we would expect the IMBHs in gas-poor non-nucleated dwarfs to be undermassive, because growth via gas accretion or the merger would be ineffective. 
Quantifying such population would offer an indirect way to test our results.

% However, there might still be additional stellar dynamical processes, not accounted for in this study, that could aid the merging of the stalled IMBHs even in non-nucleated gas-poor dwarf ellipticals.
%The inefficiency of IMBHs to merge in these galaxies also points towards a scenario where IMBHs can form a triple or quadruple system occasionally, though such system would not be stable inevitably leading to the ejection of less massive component taking energy from the tighter and massive binary's orbit \citep{man22,hau23}. In principle, such interactions can propel IMBHs binary system towards coalescence. however, the impact of such triple interaction in IMBH regime is not well explored, hence we postpone it for a subsequent study. Additionally, rotation of the IMBHs host system can have some profound effects on binary evolution, as explored in earlier studies of SMBH binary evolution in rotating environments \citep{holley+15,ras17,Mirza+17,kha20}. In the context of IMBH binary merger timescales, rotation can assist either through increase in hardening rates for IMBH binaries corotating with host sense of rotation or through increase in eccentricity for counter-rotating IMBH binaries with respect to the sense of rotation of the host \citep{sesana11,holley+15,Mirza+17}. 

However, there may be additional stellar dynamical processes, not accounted for in this study, that could aid the merging of the stalled IMBHs even in these non-nucleated gas-poor dwarfs.
An IMBH binary could remain for so long that a third IMBH can plummet into the center\citep{man22,hau23}. In principle, such interactions can propel IMBHs binary system towards coalescence, though we caution that triple interactions in the IMBH regime is not well explored. 

Additionally, dwarf galaxy rotation can have some profound effects on binary evolution, as explored in earlier studies of SMBH binary evolution in rotating environments \citep{holley+15,ras17,Mirza+17,kha20}. In the context of IMBH binary merger timescales, rotation can assist either by increasing hardening rates for co-rotating IMBH binaries or by increasing  eccentricity for counter-rotating IMBH binaries. For IMBH binaries in models under discussion, both these mechanisms might not contribute to coalescence. Co-rotation can increase hardening rates roughly by upto $30 \%$ \citep{holley+15} which is still far smaller than the values required to achieve merger in our set of simulations. Moreover, IMBH binaries evolving in co-rotation with their hosts maintain low eccentricities, resulting in a further delay to their merger time. Conversely, counter-rotation may enhance the eccentricities of intermediate-mass black hole (IMBH) binaries. However, as we notice that the eccentricities are already high for most of the models, this effect may not significantly alter findings of this study. Additionally, this enhancement is accompanied by lower hardening rates associated with binaries evolving in counter-rotation with respect to host galaxy's sense of rotation.

Another mechanism that can assist IMBHs to merge faster is the infall of globular clusters. Infalling globular clusters can add more mass to the centers and hence provide additional pool of stars for IMBH binary to interact with. 
%{ \bf Note that stars in the infalling GCs, in presence of an MBH binary in the nucleus, might be scattered via 3-body encounters, which, while they will aid the binary sinking, they would also prevent the entire stellar content of the cluster as a whole from 
%reaching the center. This could suppress the formation of an NSC compared to the case with no MBH binary present, leaving an host
%with structural properties still comparable with those of our simulated galaxies.}

%{ \bf Lucio; but wouldn't the infalling GCs form a NSC then? For dwarf galaxies I believe NSC formation through infalling GCs is the dominant channel
%in theoretical models, see papers by Niemeyer, Schoedel etc..} \textcolor{blue}{Fazeel: Yes GCs infall is a dominant model for NSC formation but the ideas here was that you have an IMBH binary that formed before the GCs reach the center; what would be impact of such an infall? It is exected that mass brought in by GCs should support stellar hardening but on the contrary, the binary would scatter the infall mass by 3-body scattering which may go against NSc fomation. This is an interesting scenario and with my MS students we are doing some initial tests on that.}
%However, it is unclear whether or not a significant mass reach the center and timescales are unclear. Currently, we are exploring this scenario and findings will be published in a future series of work.   
 Finally, even in gas-poor systems, there might be some residual gas in the center, which could arrange into a circumbinary disk surrounding the
IMBHs once their separation falls well below a parsec. This scenario has been explored extensively with numerical simulations for supermassive black hole pairs 
 (e.g. \citep{duffell20,franchini22,siwek23}),
but has not been simulated in this lower-mass regime. Using a semi-analytic framework to model the concurrent effect of gas-driven torques and stellar hardening, \citet{bortolas21} have shown,  that the combination of both mechanisms might be the key to ensuring coalescence
when the individual timescales of the two mechanisms, taken separately, are too long. Fully self-consistent, astrophysically-motivated simulations of IMBH binary formation and evolution at parsec scale are needed to properly understand the interplay between gas and stellar hardening within realistic models, and although both the hydrodynamic and stellar dynamic simulations are separately beginning to approximate a combined approach, these simulations remain an important future goal.

%%%%%%%%
%\begin{figure*}
%\centerline{
%  \resizebox{0.95\hsize}{!}{\includegraphics[angle=270]{ngc205bh.ps}}
%  }
%\caption{
%Same as those for figure \ref{fig:m32multi}, this time for NGC 205. 
%} \label{fig:ngc205param}
%\end{figure*}
%%%%%%%%
%

%%%%%%%%
%\begin{figure}
%\centerline{
%  \resizebox{0.95\hsize}{!}{\includegraphics[angle=270]{fig/ngc205estA.ps}}
%  }
%\centerline{
%  \resizebox{0.95\hsize}{!}{\includegraphics[angle=270]{fig/ngc205estB.ps}}
%  }
%\caption{
%semi-major axis (top panel) and eccentricitity (bottom panel) evoultion. 
%} \label{fig:ngc205paramnew}
%\end{figure}
%%%%%%%%

%\textbf{A few concluding remarks from Kelly }    

%%%%%%%%
%\begin{figure*}
%\centerline{
%  \resizebox{0.95\hsize}{!}{\includegraphics[angle=270]{ngc404bh.ps}}
%  }
%\caption{
%Same as those for figure \ref{fig:m32multi}, this time for NGC404. 
%} \label{fig:ngc404param}
%\end{figure*}

%%%%%%%%
%\begin{figure}
%\centerline{
%  \resizebox{0.95\hsize}{!}{\includegraphics[angle=270]{fig/ngc404estA.ps}}
%  }
%\centerline{
%  \resizebox{0.95\hsize}{!}{\includegraphics[angle=270]{fig/ngc404estB.ps}}
%  }
%\caption{
%semi-major axis (top panel) and eccentricitity (bottom panel) evoultion. 
%} \label{fig:404estimate}
%\end{figure}
%%%%%%%%

\section*{Acknowledgments}
    
We acknowledge the support by Vanderbilt University for providing access to its Advanced Computing Center for Research and Education (ACCRE). The authors also acknowledge support from High Performance Computing resources at New York University Abu Dhabi. FMK and KHB were supported through NASA ATP Grant 80NSSC18K0523. PB thanks the support from the special program of the Polish Academy of Sciences and the U.S. National Academy of Sciences under the Long-term program to support Ukrainian research teams grant No.~PAN.BFB.S.BWZ.329.022.2023.
    
\section*{Data Availability Statement}
The data underlying this article will be shared on reasonable request to the corresponding author.

\bibliography{sample631}{}
\bibliographystyle{aasjournal}



\end{document}